\documentclass[prd,superscriptaddress,twocolumn,showpacs,amsmath]{revtex4-1}
\usepackage{graphicx}
\usepackage{color}
\usepackage{booktabs}
\usepackage{hyperref}

\begin{document}
  \title{Grassmann Tensor Renormalization Group Approach to One-Flavor Lattice Schwinger Model}

  \author{Yuya Shimizu}
  \affiliation{RIKEN Advanced Institute for Computational Science, Kobe, Hyogo 650-0047, Japan}

  \author{Yoshinobu Kuramashi}
  \affiliation{Faculty of Pure and Applied Sciences, University of Tsukuba, Tsukuba, Ibaraki
    305-8571, Japan}
  \affiliation{Center for Computational Sciences, University of Tsukuba, Tsukuba, Ibaraki
    305-8577, Japan}
  \affiliation{RIKEN Advanced Institute for Computational Science, Kobe, Hyogo 650-0047, Japan}

  \begin{abstract}
    We apply the Grassmann tensor renormalization group to the lattice regularized
    Schwinger model with one-flavor of the Wilson fermion. We study the phase diagram in the $(\beta,\kappa)$ plane performing a detailed analysis of the scaling behavior of the Lee-Yang zeros and the peak height of
    the chiral susceptibility.
    Our results strongly indicate that the whole range of the phase transition line starting from $(\beta,\kappa)=(0.0,0.380665(59))$ and ending at $(\infty,0.25)$ belongs to the two-dimensional Ising universality class
    similarly to the free fermion case.
  \end{abstract}
  \date{\today}
  \maketitle

  \section{Introduction}
  \label{sec:intro}
  The Schwinger model, two-dimensional QED, has been used as a theoretical test bed for QCD. It can 
  be analytically solvable in the massless limit and has many QCD-like properties: confinement for fermions
  , chiral symmetry breaking due to the U$_A(1)$ anomaly, etc.
  The lattice regularized version of the Schwinger model is also favorable for the development of numerical
  techniques to tackle lattice QCD. The hybrid Monte Carlo algorithm (HMC) is the most successful method to
  implement dynamical fermions so far. However, it loses its validity when the determinant of the Dirac matrix can
  be negative. Such a difficulty has been preventing us from studying the phase structure of the one-flavor lattice Schwinger
  model in the Wilson fermion formulation. 
  A system of free Wilson fermions at $\beta=\infty$ exhibits a second order phase transition at $\kappa=0.25$ with $\kappa$
  the hopping parameter. It belongs to the 2d Ising universality class. In the strong coupling limit at $\beta=0.0$, the
  one-flavor lattice Schwinger model was shown to be mapped to an eight-vertex model~\cite{Salmhofer:1991cc}. This was followed by a large
  scale Monte Carlo simulation on spherelike lattices, which has proved that this model also lies in the 2d Ising
  universality class~\cite{Gausterer:1995np}. One may naively expect that a phase transition line runs from $\beta=0.0$ to $\infty$ belonging to the 2d Ising universality class. A result
  obtained by the microcanonical fermionic average approach, however, indicates that the phase transition at finite
  $\beta$ lies in a different universality class from the 2d Ising model~\cite{Azcoiti:1995bx}.
  Furthermore, an analysis of the weak coupling expansion on large lattices disproves even the existence
  of the phase transition line~\cite{Kenna:1998rs}.

  The tensor
  network renormalization group (TRG) was originally introduced by Levin and Nave~\cite{Levin:2007aa}. 
It has been applied to a couple of models consisting of continuous bosonic
variables~\cite{Shimizu:2012zza,Liu:2013nsa,Meurice:2013cla,
  Yu:2013aa,Denbleyker:2013bea}. Gu {\it et al.} generalized the TRG to a Grassmann valued tensor network in
  order to investigate fermionic systems~\cite{Gu:2010aa,Gu:2011aa}.
  Although direct evaluation of multiple Grassmann integrals is an exponentially hard task as discussed by
  Creutz~\cite{Creutz:1998ee}, the Grassmann TRG (GTRG) allows us to evaluate the
  partition function and expectation values of physical quantities with reasonable amount of computational resources even for fermionic
  systems. In this paper, we apply the GTRG to the lattice Schwinger model with one-flavor of the Wilson fermion.
  We demonstrate that the GTRG works well even at the critical hopping parameter where the negative sign from
  the fermion determinant may arise, and determines the phase structure of the one-flavor lattice Schwinger
  model.

  We mention that there are some related works with different approaches, where the density matrix renormalization group or
  variational matrix-product-state method are employed\cite{Byrnes:2002nv,Banuls:2013jaa,Banuls:2013zva,Buyens:2013yza,Riko:2013aa}.
  They are all based on the Hamiltonian lattice
  gauge theory with the Kogut-Susskind formulation. 

  This paper is organized as follows. In Sec.~\ref{sec:gtrg}, we explain the GTRG procedure 
for the tensor network showing its
  representation of the partition function of the lattice Schwinger model.
  We present numerical results for the finite size scaling analyses in Sec.~\ref{sec:num}.
  Sec.~\ref{sec:sum} is devoted to summary and outlook.

  \section{Grassmann Tensor Renormalization Group for the Lattice Schwinger Model}
  \label{sec:gtrg}
  \subsection{Lattice formulation}
  The partition function of lattice gauge theory can be generally expressed as
  \begin{equation}
    Z=\int\!\mathcal{D}U\,\det D[U]\,e^{-S_g[U]}, 
  \end{equation}
  where $S_g$ is the gauge action and its expression will be given below.
  We employ the Wilson fermion formulation, whose Dirac matrix $D[U]$ is given by
  \begin{equation}
    \begin{split}
      \Bar{\psi}D[U]\psi=&\frac{1}{2\kappa}\sum_{n,\alpha} \Bar{\psi}_{n,\alpha}\psi_{n,\alpha}\\
      &-\frac{1}{2}\sum_{n,\mu,\alpha,\beta}\Bar{\psi}_{n,\alpha}\{
      (1-\gamma_\mu)_{\alpha,\beta}\,U_{n,\mu}\psi_{n+\Hat{\mu},\beta}\\&+(1+\gamma_\mu)_{\alpha,\beta}
      \,U^\dagger_{n-\Hat{\mu},\mu}\psi_{n-\Hat{\mu},\beta}\},\\
      \equiv& S_f[\psi,\Bar{\psi},U]
    \end{split}
  \end{equation}
  with $\kappa$ the hopping parameter and $U_{n,\mu}$ an $U(1)$ link variable at site $n$ along $\mu$
  direction. $\alpha,\beta$ denote the Dirac indices and
$\Hat{\mu}$ represents an unit vector along $\mu$ direction. 
  The Grassmann path integral representation for $\det D[U]$ is given by
  \begin{equation}
    \det D[U]=\prod_{n,\alpha}\left(\int\!d\psi_{n,\alpha} d\Bar{\psi}_{n,\alpha}\right)
    e^{S_f[\psi,\Bar{\psi},U]},\label{eq:detd}
  \end{equation}
  where the Grassmann variables $\{\psi_{n,\alpha}\}$ and $\{\Bar{\psi}_{n,\alpha}\}$ satisfy the following relations:
  \begin{gather}
    [\psi_{n,\alpha},\Bar{\psi}_{m,\beta}]_+\equiv \psi_{n,\alpha}\Bar{\psi}_{m,\beta}
    +\Bar{\psi}_{m,\beta}\psi_{n,\alpha}=0,\\
    [\psi_{n,\alpha},\psi_{m,\beta}]_+=[\Bar{\psi}_{n,\alpha},\Bar{\psi}_{m,\beta}]_+=0,\\
    \int\!d\psi_{n,\alpha}\,1=\int\!d\Bar{\psi}_{n,\alpha}\,1=0,\\
    \int\!d\psi_{n,\alpha}\,\psi_{m,\beta}=\int\!d\Bar{\psi}_{n,\alpha}\,\Bar{\psi}_{m,\beta}=\delta_{n,m}
    \delta_{\alpha,\beta}.
  \end{gather}
  With the choice of a representation of gamma matrices 
  \begin{equation}
    \gamma_1=\sigma_3=
    \begin{pmatrix}
      1&0\\
      0&-1
    \end{pmatrix},\quad
    \gamma_2=\sigma_1=
    \begin{pmatrix}
      0&1\\
      1&0
    \end{pmatrix},
  \end{equation}
  we introduce another basis:
  \begin{gather}
    \chi_{n,1}=\frac{1}{\sqrt{2}}(\psi_{n,1}+\psi_{n,2}),\ 
    \chi_{n,2}=\frac{1}{\sqrt{2}}(\psi_{n,1}-\psi_{n,2}),\\
    \Bar{\chi}_{n,1}=\frac{1}{\sqrt{2}}(\Bar{\psi}_{n,1}+\Bar{\psi}_{n,2}),\ 
    \Bar{\chi}_{n,2}=\frac{1}{\sqrt{2}}(\Bar{\psi}_{n,1}-\Bar{\psi}_{n,2}),
  \end{gather}
  which yields
  \begin{gather}
    \sum_{\alpha,\beta}\Bar{\psi}_{n,\alpha}(1+\gamma_1)_{\alpha,\beta}\psi_{n-\Hat{1},\beta}
    =2\Bar{\psi}_{n,1}\psi_{n-\Hat{1},1},\\
    \sum_{\alpha,\beta}\Bar{\psi}_{n,\alpha}(1-\gamma_1)_{\alpha,\beta}\psi_{n+\Hat{1},\beta}
    =2\Bar{\psi}_{n,2}\psi_{n+\Hat{1},2},\\
    \sum_{\alpha,\beta}\Bar{\psi}_{n,\alpha}(1+\gamma_2)_{\alpha,\beta}\psi_{n-\Hat{2},\beta}
    =2\Bar{\chi}_{n,1}\chi_{n-\Hat{2},1},\\
    \sum_{\alpha,\beta}\Bar{\psi}_{n,\alpha}(1-\gamma_2)_{\alpha,\beta}\psi_{n+\Hat{2},\beta}
    =2\Bar{\chi}_{n,2}\chi_{n+\Hat{2},2}.
  \end{gather}
  Notice that $\{\chi_{n,\alpha}\}$ and $\{\Bar{\chi}_{n,\alpha}\}$ also satisfy anticommutation relations:
  \begin{gather}
    [\chi_{n,\alpha},\Bar{\chi}_{m,\beta}]_+=[\chi_{n,\alpha},\chi_{m,\beta}]_+
    =[\Bar{\chi}_{n,\alpha},\Bar{\chi}_{m,\beta}]_+=0.
  \end{gather}

  \subsection{Grassmann valued tensor network}
  \label{sec:gtn}
  We first transform $\det D[U]$ into a tensor network.
  The exponential form in Eq.~\eqref{eq:detd} is expanded as follows by using the anticommutation property of the   Grassmann variables:
  \begin{equation}
  \begin{split}
    e^{S_f[\psi,\Bar{\psi},U]}=&\prod_{n}\left(1+\frac{1}{2\kappa}\Bar{\psi}_{n,1}\psi_{n,1}\right)
    \left(1+\frac{1}{2\kappa}\bar{\psi}_{n,2}\psi_{n,2}\right)\\
    &\times\left(1-U^\dagger_{n-\Hat{1},1}\Bar{\psi}_{n,1}\psi_{n-\Hat{1},1}\right)\\
    &\times\left(1-U_{n,1}\Bar{\psi}_{n,2}\psi_{n+\Hat{1},2}\right)\\
    &\times\left(1-U^\dagger_{n-\Hat{2},2}\Bar{\chi}_{n,1}\chi_{n-\Hat{2},1}\right)\\
    &\times\left(1-U_{n,2}\Bar{\chi}_{n,2}\chi_{n+\Hat{2},2}\right).
  \end{split}
  \end{equation}
  Here we define the Wilson term and the hopping terms as
  \begin{gather}
  \begin{split}
    W_n\equiv &\frac{1}{4\kappa^2}+\frac{1}{2\kappa}d\psi_{n,1}d\Bar{\psi}_{n,1}
    +\frac{1}{2\kappa}d\psi_{n,2}d\Bar{\psi}_{n,2}\\
    &+d\psi_{n,1}d\Bar{\psi}_{n,1}d\psi_{n,2}d\Bar{\psi}_{n,2},
  \end{split}\\
  \begin{split}
    \Tilde{H}_{n,1}\equiv&1-U^\dagger_{n,1}\Bar{\psi}_{n+\Hat{1},1}\psi_{n,1}
    -U_{n,1}\Bar{\psi}_{n,2}\psi_{n+\Hat{1},2}\\&+\Bar{\psi}_{n+\Hat{1},1}\psi_{n,1}
    \Bar{\psi}_{n,2}\psi_{n+\Hat{1},2},
  \end{split}\\
  \begin{split}
    \Tilde{H}_{n,2}\equiv&1-U^\dagger_{n,2}\Bar{\chi}_{n+\Hat{2},1}\chi_{n,1}
    -U_{n,2}\Bar{\chi}_{n,2}\psi_{n+\Hat{2},2}\\&+\Bar{\chi}_{n+\Hat{2},1}\chi_{n,1}
    \Bar{\chi}_{n,2}\chi_{n+\Hat{2},2}.
  \end{split}
  \end{gather}
  The determinant $\det D[U]$ is expressed as a simple form: 
  \begin{equation}
    \det D[U]=P_0\int\prod_n W_n\prod_\mu \Tilde{H}_{n,\mu},
    \label{}
  \end{equation}
  where $P_0$ represents a projection to terms without any Grassmann variable.

\begin{figure}
    \centering
    \includegraphics[width=60mm]{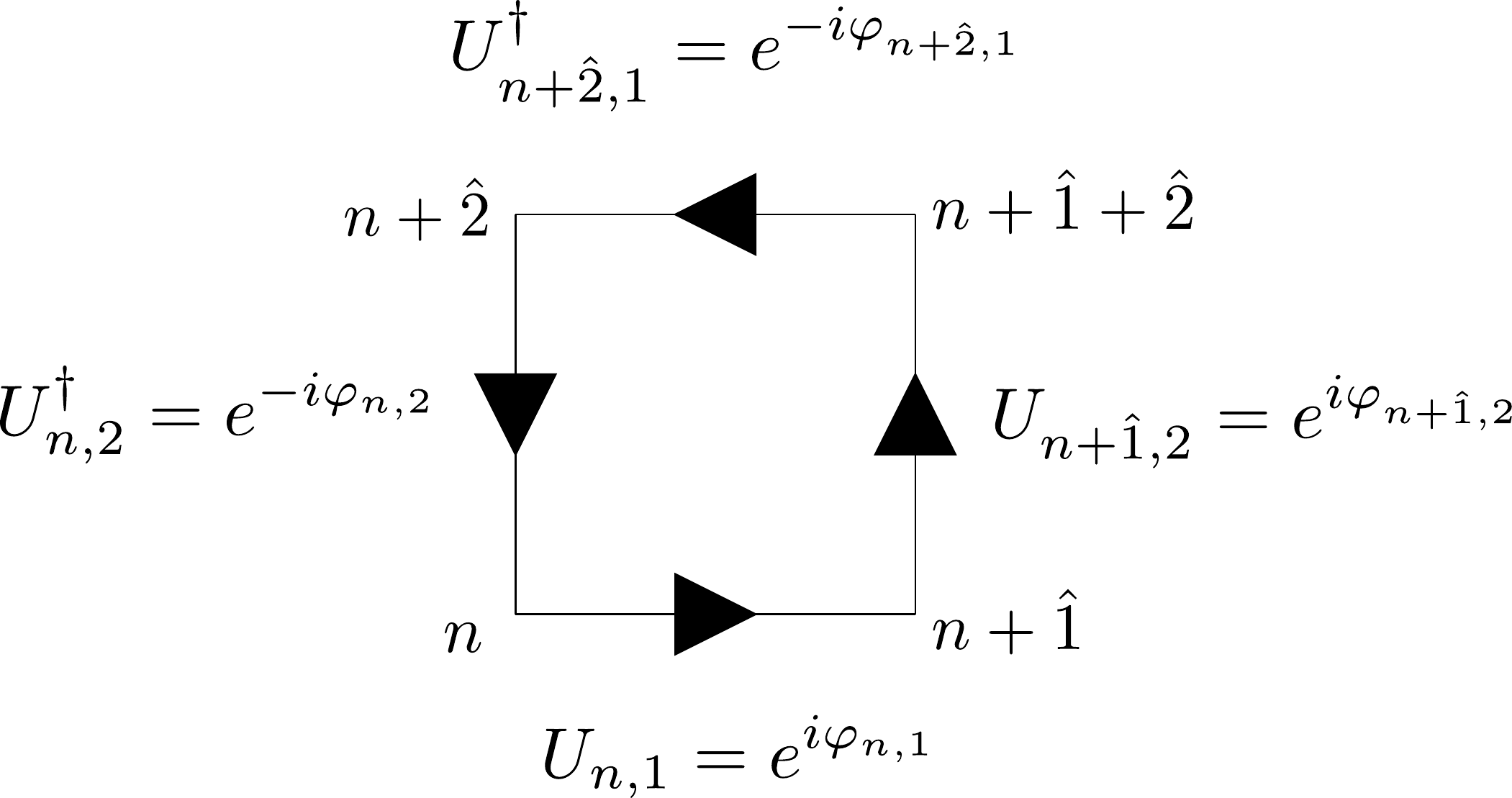}
    \caption{Plaquette as a product of link variables.}
    \label{fig:plq}
  \end{figure}

  Let us turn to the gauge part. We employ the U(1) plaquette gauge action:
  \begin{gather}
    S_g=-\beta\sum_{p}\cos\varphi_p,\\
    \varphi_p=\varphi_{n,1}+\varphi_{n+\Hat{1},2}-\varphi_{n+\Hat{2},1}-\varphi_{n,2},\\
    \varphi_{n,1},\varphi_{n+\Hat{1},2},\varphi_{n+\Hat{2},1},\varphi_{n,2}\in [-\pi,\pi],
  \end{gather}
  where $\varphi_{n,1},\varphi_{n+\Hat{1},2},\varphi_{n+\Hat{2},1}$ and $\varphi_{n,2}$ are phases of
  U(1) link variables which compose a plaquette variable $\varphi_p$ depicted in Fig.~\ref{fig:plq}.
  $\beta$ is the inverse coupling constant squared.
  Liu {\it et al.} have shown that a finite-dimensional tensor network representation of pure lattice gauge
  theory is derived by the character expansion (CE) with truncation~\cite{Liu:2013nsa}, and its numerical accuracy with the
  TRG is verified for the $XY$ model~\cite{Meurice:2013cla,Yu:2013aa,Denbleyker:2013bea}.
  Using the character expansion, the Boltzmann weight per plaquette is decomposed as
  \begin{align}
    e^{\beta\cos\varphi_p}&=\sum_{m_b=-\infty}^\infty e^{im_b\varphi_p}I_{m_b}(\beta),\\
    &\simeq\sum_{m_b=N_{\rm ce}}^{N_{\rm ce}} e^{im_b\varphi_p}I_{m_b}(\beta),
  \end{align}
  where $I_{m_b}$ is the modified Bessel function and $N_{\rm ce}$ is the truncation number in the character expansion. The subscript $b$ denotes bosonic indices.
  After integrating out all the link variables, the hopping term is written as
  \begin{equation}
    \begin{split}
      H_{n,1;m_b,n_b}&\equiv
      \int_{-\pi}^{\pi}\!\frac{d\varphi_{n,1}}{2\pi}\Tilde{H}_{n,1}\,e^{i(m_b-n_b)\varphi_{n,1}},\\
      &=
      \begin{cases}
        1+\Bar{\psi}_{n+\Hat{1},1}\psi_{n,1}\Bar{\psi}_{n,2}\psi_{n+\Hat{1},2}&m_b=n_b\\
        -\Bar{\psi}_{n+\Hat{1},1}\psi_{n,1}&m_b=n_b+1\\
        -\Bar{\psi}_{n,2}\psi_{n+\Hat{1},2}&m_b=n_b-1\\
        0&\text{others}
      \end{cases}.
    \end{split}
  \end{equation}
  $H_{n,2;m_b,n_b}$ is given in the same manner.

  Now we introduce a tensor form of $W_n$ and $H_{n,\mu;m_b,n_b}$:
  \begin{gather}
    \begin{split}
      W_n=\sum_{i_{f1},i_{f2},\cdots=0}^1&W^{i_{f1},i_{f2},j_{f1},j_{f2},k_{f1},k_{f2},l_{f1},l_{f2}}\\
      &d\Bar{\psi}_{n,2}^{i_{f2}}\,d\psi_{n,1}^{i_{f1}}\,d\Bar{\chi}_{n,2}^{j_{f2}}\,d\chi_{n,1}^{j_{f1}}\,
      d\psi_{n,2}^{k_{f2}}\,d\Bar{\psi}_{n,1}^{k_{f1}}\\&d\chi_{n,2}^{l_{f2}}\,d\Bar{\chi}_{n,1}^{l_{f1}},
    \end{split}\\
      H_{n,1;m_b,n_b}=\sum_{i_{f1},i_{f2}=0}^1H_{m_b,n_b}^{i_{f1},i_{f2}}\,\Bar{\psi}_{n+\Hat{1},1}^{i_{f1}}
      \psi_{n+\Hat{1},2}^{i_{f2}}\psi_{n,1}^{i_{f1}}\Bar{\psi}_{n,2}^{i_{f2}},\\
      H_{n,2;m_b,n_b}=\sum_{i_{f1},i_{f2}=0}^1H_{m_b,n_b}^{i_{f1},i_{f2}}\,\Bar{\chi}_{n+\Hat{2},1}^{i_{f1}}
      \chi_{n+\Hat{2},2}^{i_{f2}}\chi_{n,1}^{i_{f1}}\Bar{\chi}_{n,2}^{i_{f2}},
  \end{gather}
  where ${f1},{f2}$ denote fermionic indices.
  Each value for $W^{i_{f1},\cdots,l_{f2}}$ and $H_{m_b,n_b}^{i_{f1},i_{f2}}$ is given in Appendix~\ref{sec:appa}.
 
  Finally, the partition function $Z$ is written as a tensor network form
  \begin{equation}
    Z=\int\sum_{i,j,k,\cdots}T_{n;i,j,k,l}\,T_{n+\Hat{1};m,o,i,p}\,T_{n+\Hat{2};q,r,s,j}\,\cdots,
    \label{}
  \end{equation}
where $i,\,j,\,k,\,\cdots$ are combinations of bosonic and fermionic indices, namely,
 $i=(i_b,i_{f1},i_{f2})$ and the tensors are expressed as
  \begin{gather}
    \begin{split}
      T_{n;i,j,k,l}\equiv &T_{i,j,k,l}\,d\Bar{\psi}_{n,2}^{i_{f2}}\,d\psi_{n,1}^{i_{f1}}
      \,d\Bar{\chi}_{n,2}^{j_{f2}}\,d\chi_{n,1}^{j_{f1}}\,d\psi_{n,2}^{k_{f2}}\\&d\Bar{\psi}_{n,1}^{k_{f1}}
      \,d\chi_{n,2}^{l_{f2}}\,d\Bar{\chi}_{n,1}^{l_{f1}}\,\Bar{\psi}_{n+\Hat{1},1}^{i_{f1}}
      \psi_{n+\Hat{1},2}^{i_{f2}}\psi_{n,1}^{i_{f1}}\Bar{\psi}_{n,2}^{i_{f2}}\\
      &\Bar{\chi}_{n+\Hat{2},1}^{j_{f1}}\chi_{n+\Hat{2},2}^{j_{f2}}
      \chi_{n,1}^{j_{f1}}\Bar{\chi}_{n,2}^{j_{f2}}
    \end{split}
\end{gather}
with
\begin{gather}
    \begin{split}
      T_{i,j,k,l}\equiv&W^{i_{f1},i_{f2},j_{f1},j_{f2},k_{f1},k_{f2},l_{f1},l_{f2}}\\
      &H_{k_b,i_b}^{i_{f1},i_{f2}}\,H_{i_b,l_b}^{j_{f1},j_{f2}}\,I_{i_b}(\beta)\,\delta_{i_b,j_b}.
    \end{split}
  \end{gather}
Note that we treat $\psi_{n,\alpha},\,\chi_{n,\alpha}$, and others also, as independent variables.

  \subsection{Grassmann tensor renormalization group}
  We define a Grassmann version of the singular value decomposition (SVD) in a similar manner as in Ref.~\cite{Gu:2010aa,Gu:2011aa}:
  \begin{gather}
    T_{n_e;i,j,k,l}=\sum_{p,q}\int S^{1}_{n^\prime+\Hat{1};i,j,p}\,S^{3}_{n^\prime;k,l,q}
    \,g^{13}_{n^\prime;p,q}
\end{gather}
with
\begin{gather}
    \begin{split}
    g^{13}_{n^\prime;p,q}\equiv \Bar{\xi}_{n^\prime+\Hat{1}}^{p_f}\xi_{n^\prime}^{q_f}\,\delta_{p_b,q_b},
    \end{split}\\
    \begin{split}
    S^{1}_{n^\prime+\Hat{1};i,j,p}\equiv&S^{1}_{i,j,p}\,d\Bar{\xi}^{p_f}_{n^\prime+\Hat{1}}
    \,d\Bar{\psi}_{n_e,2}^{i_{f2}}\,d\psi_{n_e,1}^{i_{f1}}\,d\Bar{\chi}_{n_e,2}^{j_{f2}}\\
    &d\chi_{n_e,1}^{j_{f1}}\,\Bar{\psi}_{n_e+\Hat{1},1}^{i_{f1}}\psi_{n_e+\Hat{1},2}^{i_{f2}}
    \psi_{n_e,1}^{i_{f1}}\Bar{\psi}_{n_e,2}^{i_{f2}}\\
    &\Bar{\chi}_{n_e+\Hat{2},1}^{j_{f1}}\chi_{n_e+\Hat{2},2}^{j_{f2}}\chi_{n_e,1}^{j_{f1}}
    \Bar{\chi}_{n_e,2}^{j_{f2}},
    \end{split}\\
    \begin{split}
      S^{3}_{n^\prime;k,l,q}\equiv&S^{3}_{k,l,q}\,d\xi^{q_f}_{n^\prime}\,d\psi_{n_e,2}^{k_{f2}}
      \,d\Bar{\psi}_{n_e,1}^{k_{f1}}\\&d\chi_{n_e,2}^{l_{f2}}\,d\Bar{\chi}_{n_e,1}^{l_{f1}},
    \end{split}
  \end{gather}
and
  \begin{gather}
    T_{n_o;i,j,k,l}=\sum_{r,s}\int S^{2}_{n^\prime+\Hat{2};l,i,r}\,S^{4}_{n^\prime;j,k,s}
    \,g^{24}_{n^\prime;r,s}
  \end{gather}
with
\begin{gather}
    \begin{split}
    g^{24}_{n^\prime;r,s}\equiv& \Bar{\eta}_{n^\prime+\Hat{2}}^{r_f}\eta_{n^\prime}^{s_f}\,\delta_{r_b,s_b},
    \end{split}\\
    \begin{split}
      S^{2}_{n^\prime+\Hat{2};l,i,r}\equiv& S^{2}_{l,i,r}\,d\Bar{\eta}^{r_f}_{n^\prime+\Hat{2}}
      \,d\chi_{n_o,2}^{l_{f2}}
      \,d\Bar{\chi}_{n_o,1}^{l_{f1}}\,d\Bar{\phi}_{n_o,2}^{i_{f2}}\\&d\phi_{n_o,1}^{i_{f1}}
      \,\Bar{\psi}_{n_o+\Hat{1},1}^{i_{f1}}\psi_{n_o+\Hat{1},2}^{i_{f2}}\psi_{n_o,1}^{i_{f1}}
      \Bar{\psi}_{n_o,2}^{i_{f2}},
    \end{split}\\
    \begin{split}
      S^{4}_{n^\prime;j,k,s}\equiv& S^{4}_{j,k,s}\,d\eta_{n^\prime}^{s_f}
      \,d\Bar{\chi}_{n_o,2}^{j_{f2}}\,d\chi_{n_o,1}^{j_{f1}}
      \,d\phi_{n_o,2}^{k_{f2}}\\&d\Bar{\phi}_{n_o,1}^{k_{f1}}\,\Bar{\chi}_{n_o+\Hat{2},1}^{j_{f1}}
      \chi_{n_o+\Hat{2},2}^{j_{f2}}\chi_{n_o,1}^{j_{f1}}\Bar{\chi}_{n_o,2}^{j_{f2}},
    \end{split}
  \end{gather}
  where $\Bar{\xi}_{n^\prime},\,\xi_{n^\prime},\,\Bar{\eta}_{n^\prime},\,\eta_{n^\prime}$ are Grassmann
  variables with $n^\prime$ the coarse-grained lattice site.
  $n_e$ and $n_o$ mean even and odd sites respectively. $S^{1}_{i,j,p}$ and $S^{3}_{k,l,q}$ are
  determined by the SVD for the matrix $M^{13}_{(i,j),(k,l)}\equiv T_{i,j,k,l}$:
  \begin{gather}
    M^{13}_{(i,j),(k,l)}=\sum_m U_{(i,j),m} \sigma_m V_{(k,l),m},\\
    S^1_{i,j,p}=\sqrt{\sigma_p}\,U_{(i,j),p},\\
    S^3_{k,l,q}=\sqrt{\sigma_q}\,V_{(k,l),q}.
  \end{gather}
  In numerical calculation, we keep only the largest $D$ singular values out of $\{\sigma_m\}$.
  $S^{2}_{j,k,r}$ and $S^{4}_{l,i,s}$ are determined similarly for the
  matrix $M^{24}_{(l,i),(j,k)}\equiv (-1)^{i_{f1}+i_{f2}}T_{i,j,k,l}$.
  Although the dimension of $M^{13}$ and $M^{24}$ is $16\times(2N_{\rm ce}+1)^2$, the calculational cost of the
  SVD can be drastically reduced because of the following conditions:
  \begin{gather}
    \begin{split}
    &i_{f1}+i_{f2}+j_{f1}+j_{f2}\\
    &+k_{f1}+k_{f2}+l_{f1}+l_{f2}=\text{even},
    \end{split}\\
    i_b=j_b.
    \label{}
  \end{gather}
  Finally, we obtain a coarse-grained tensor $T^\prime_{n^\prime;q,r,p,s}$,
  \begin{equation}
    \begin{split}
      T^\prime_{n^\prime;q,r,p,s}&=\sum_{i,j,k,l}\int S^1_{n^\prime;i,j,p}S^2_{n^\prime;j,k,r}
      S^3_{n^\prime;k,l,q}S^4_{n^\prime;l,i,s}\\
      &
      \begin{split}
      =&\sum_{i,j,k,l}(-1)^{i_{f1}+i_{f2}}S^1_{i,j,p}S^2_{j,k,r}S^3_{k,l,q}
      S^4_{l,i,s}\\
      &d\xi^{q_f}_{n^\prime}\,d\eta^{r_f}_{n^\prime}\,d\Bar{\xi}^{p_f}_{n^\prime}
      \,d\Bar{\eta}^{s_f}_{n^\prime}.
      \end{split}
    \end{split}
    \label{}
  \end{equation}
  The partition function is reexpressed as
  \begin{equation}
    \begin{split}
      Z=&\int\sum_{i,j,k,\cdots}T^\prime_{n^\prime;i,j,k,l}\,T^\prime_{n^\prime+\Hat{1};m,o,t,p}
      \,T^\prime_{n^\prime+\Hat{2};q,r,s,u}\,\cdots\\
      &g^{13}_{n^\prime;t,i}\,g^{24}_{n^\prime;u,j}\,\cdots,
    \end{split}
    \label{}
  \end{equation}
where only new Grassmann variables
  $\Bar{\xi}_{n^\prime},\,\xi_{n^\prime},\,\Bar{\eta}_{n^\prime},\,\eta_{n^\prime}$ remain after old ones $\Bar{\psi}_{n,\mu},\,\psi_{n,\mu},\,\Bar{\chi}_{n,\mu},\,\chi_{n,\mu}$ were integrated out.
  Note that the scaling factor of this transformation is $\sqrt{2}$. Further iterative transformations can be performed in the same way.

  \section{Numerical Results}
  \label{sec:num}
  \subsection{Setup}

  \begin{table}
    \centering
    \caption{Parameters in our numerical analysis.}
    \label{tab:param}
    \begin{ruledtabular}
    \begin{tabular}{clc}\toprule
      parameter&description& value \\ \hline
      $N_{\rm ce}$& truncation number of CE & $15$ \\
      $D$& truncation number of SVD& $96$ \\
      $\beta$& inverse coupling constant squared & $0.0,5.0,10.0$ \\
      $\kappa$& hopping parameter & $\beta$ dependent \\ \bottomrule
    \end{tabular}
    \end{ruledtabular}
  \end{table}

\begin{figure}
    \centering
    \includegraphics[width=80mm]{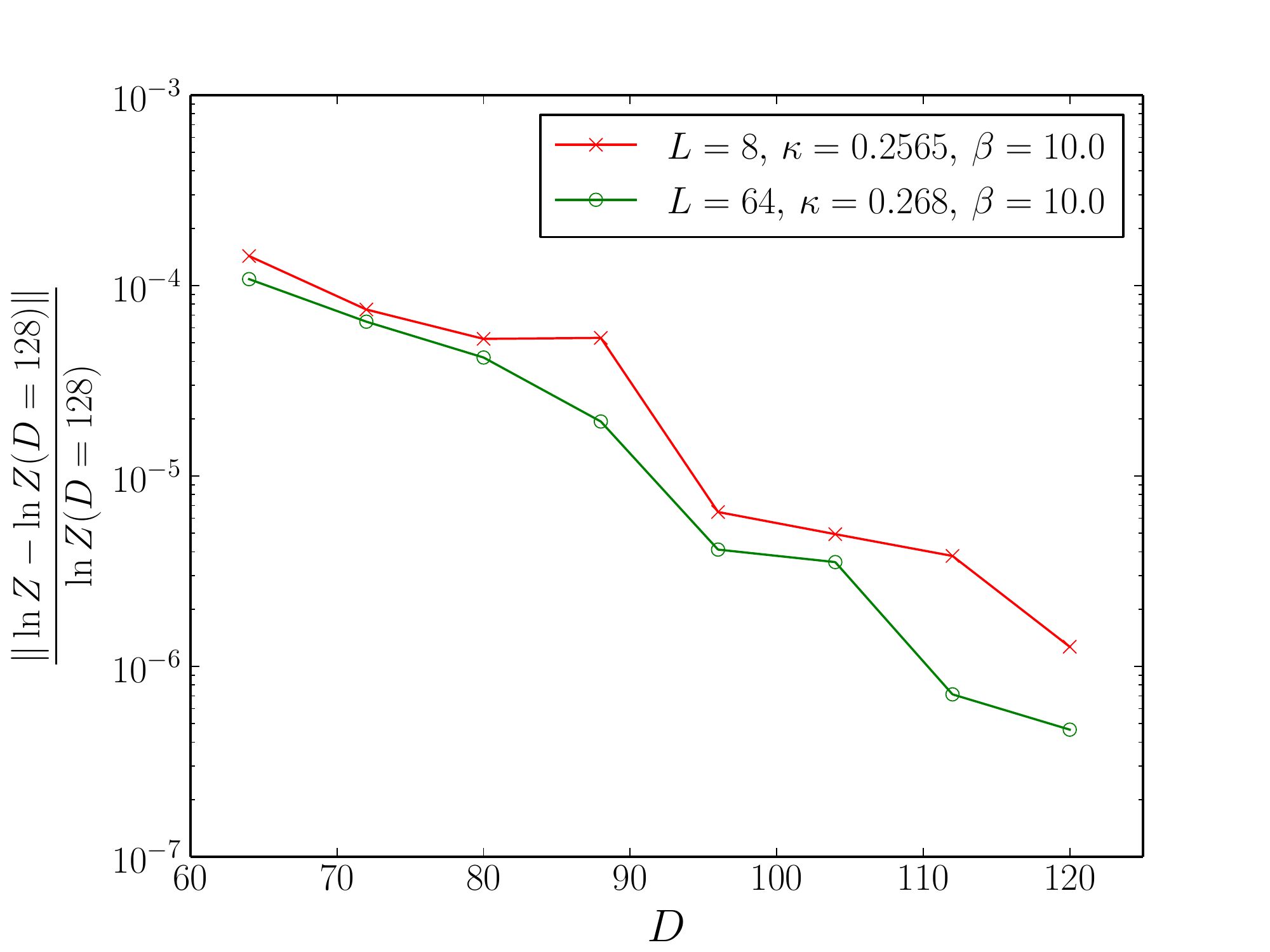}
    \caption{Convergence of $\ln Z$ as a function of SVD truncation number $D$.}
    \label{fig:err}
  \end{figure}

  We list parameters in our numerical analysis in Table~\ref{tab:param}.
  $N_{\rm ce}$ and $D$ are truncation parameters in the TRG procedure as explained in Sec.~\ref{sec:gtrg}.
  We choose $N_{\rm ce}=15$ and $D=96$, which provide us sufficiently accurate
  results for all the $\beta$ and $\kappa$ values employed in this work.
  We calculate the partition function $Z$ at the strong coupling limit ($\beta=0.0$) and finite couplings
  ($\beta=5.0,\,10.0$). Figure~\ref{fig:err} shows a typical example of convergence behavior of $\ln Z$ as a function of $D$. We find the value of $\ln Z$ with $D=64$ already reach a high precision.
  Since the scaling factor of the TRG is $\sqrt{2}$, we are allowed to evaluate physical quantities not only at the lattice size
  $L=4,\,8,\,16,\,\cdots$, but also at $L=4\sqrt{2},\,8\sqrt{2},\,16\sqrt{2},\,\cdots$. The periodic boundary
  condition is employed.
 
  We have performed two kinds of finite size scaling analyses. One is an investigation of the scaling properties of the peak height of the chiral susceptibility which is obtained by differentiating $({\ln Z})/{L^2}$ twice with respect to ${1}/{(2\kappa)}$:
  \begin{align}
    \chi(L)&\equiv \sum_n\langle\Bar{\psi}_n\psi_n\Bar{\psi}_0\psi_0\rangle-L^2\langle\Bar{\psi_0}\psi_0
    \rangle^2,\\
    &=\frac{1}{L^2}\frac{\partial^2\ln Z}{\partial (1/2\kappa)^2}.
    \label{eq:chi}
  \end{align}
  The chiral susceptibility has a peak at the critical hopping parameter $\kappa_c(L)$ where the fermion mass is expected to vanish
  and the correlation length diverges.
  The other is the so-called Lee-Yang zero analysis in the complex $\kappa$ plane. We have investigated the scaling behaviors of both the real and imaginary parts of the partition function
  zeros which approach to $\kappa_c$ in the infinite volume limit.

  \subsection{Strong coupling limit ($\beta=0.0$)}
\begin{figure}
    \centering
    \includegraphics[width=80mm]{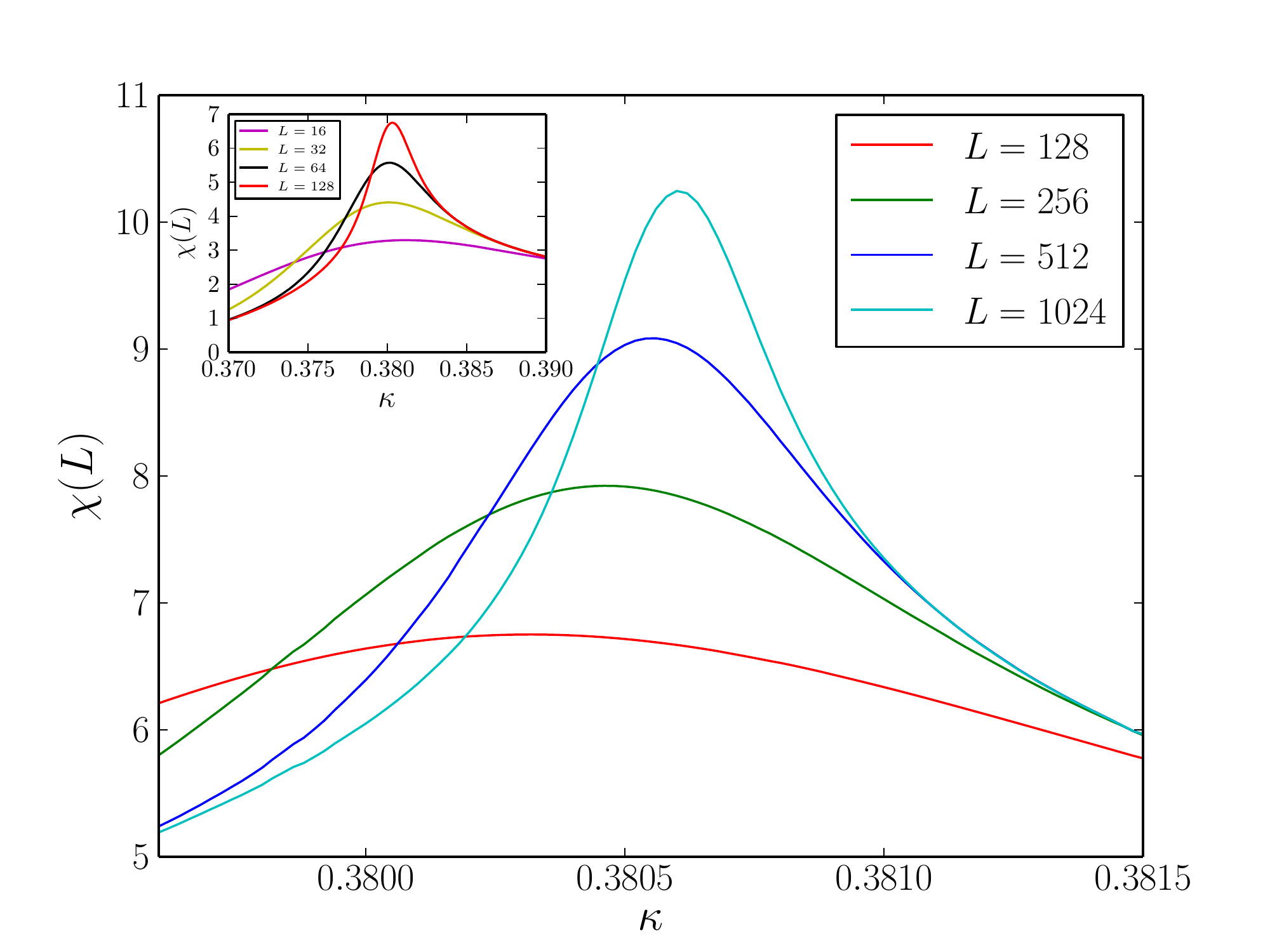}
    \caption{Chiral susceptibility $\chi(L)$ as a function of hopping parameter $\kappa$ at $\beta=0.0$ for $L=16,32,64,\dots,1024$.}
    \label{fig:chi}
  \end{figure}

\begin{figure}
    \centering
    \includegraphics[width=80mm]{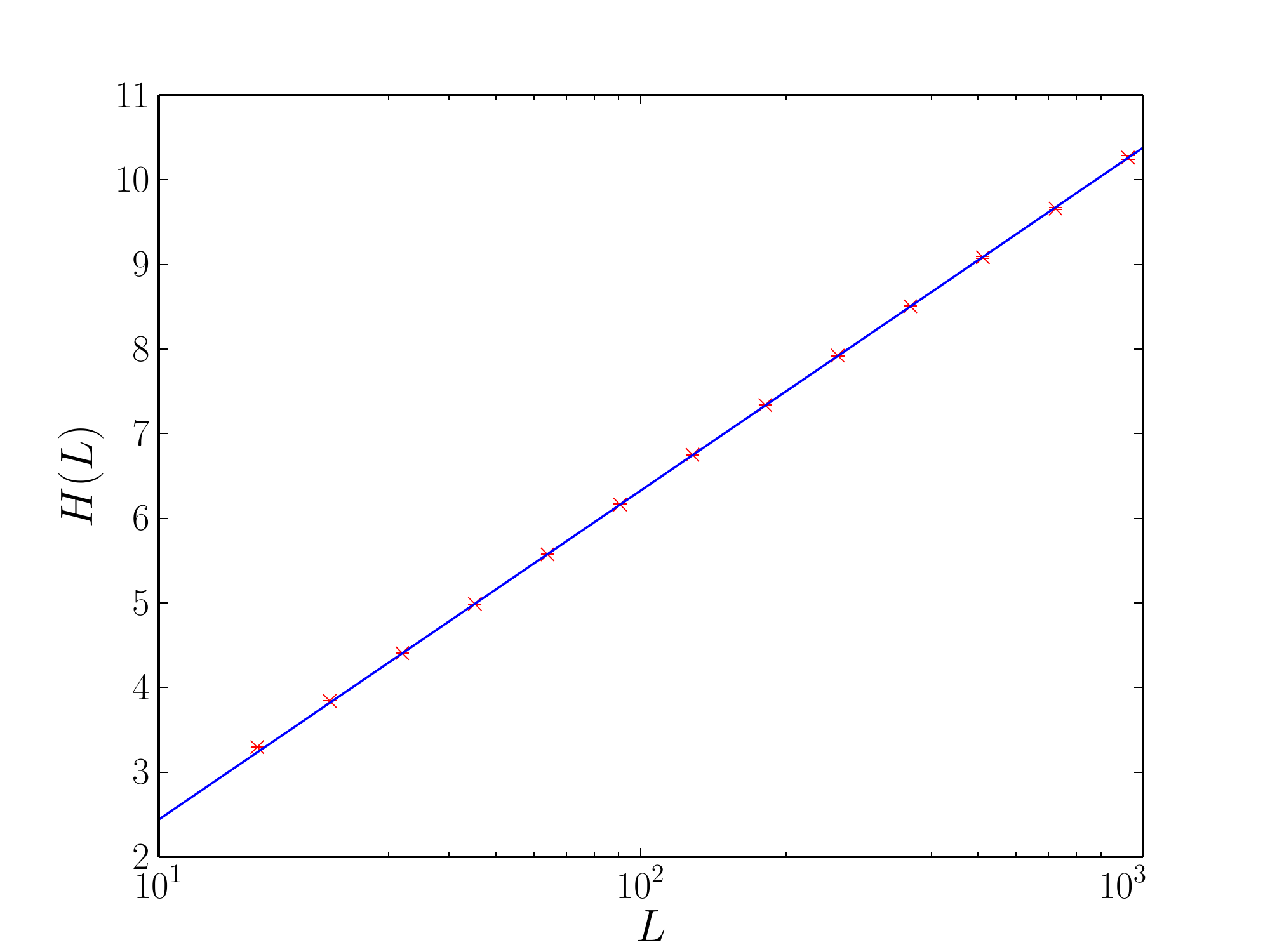}
    \caption{Peak height of the chiral susceptibility $H(L)$ as a function of $L$ at $\beta=0.0$. The horizontal axis is logarithmic. Solid line represents a linear fit in terms of $\ln Z$.}
    \label{fig:ph000}
  \end{figure}

\begin{figure}
    \centering
    \includegraphics[width=80mm]{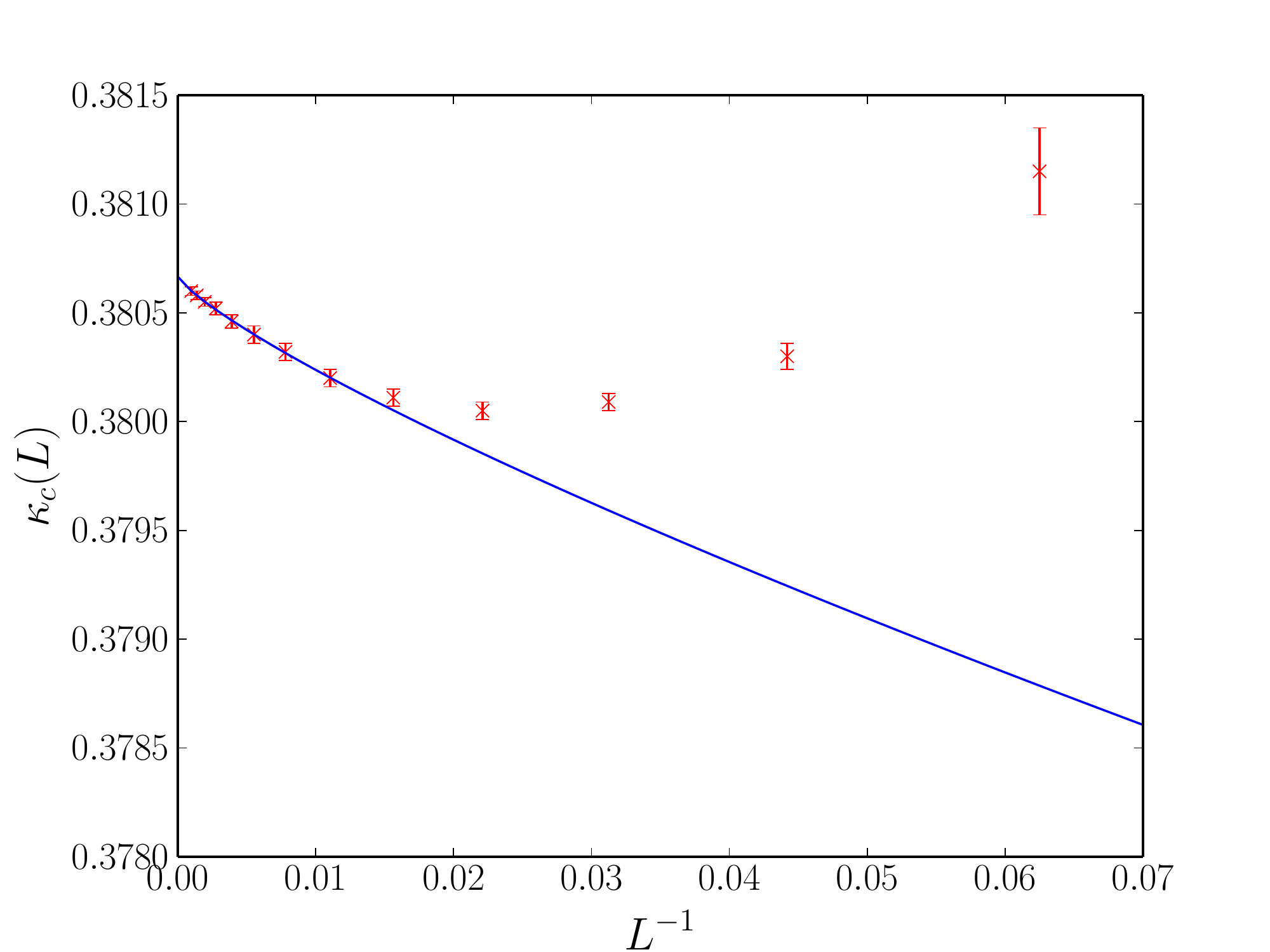}
    \caption{Peak position of the chiral susceptibility $\kappa_c(L)$ as a function of $L^{-1}$ at $\beta=0.0$. Solid curve represents the fit result.}
    \label{fig:kc000}
  \end{figure}

  The strong coupling limit is a special case. Since we are allowed to integrate out all the Grassmann
  variables analytically, we can employ the conventional TRG instead of the GTRG. This enables us to make a more precise numerical analysis at the strong coupling limit than at finite coupling.
  Figure~\ref{fig:chi} plots the chiral susceptibility as a function of $\kappa$.
  We observe the clear peak structure at all the values of $L$ and the peak height grows as $L$ increases.
  In case of the one-flavor Schwinger model, chiral symmetry is always broken because of the U$_A(1)$ anomaly even
  in the continuum limit.
  Therefore, we expect that the peak height $H(L)$ scales with $L$ as
  \begin{equation}
    H(L)\propto L^{\alpha/\nu},
    \label{}
  \end{equation}
  where $\alpha$ is the critical exponent for the heat capacity rather than that for the susceptibility.
 We plot the peak height $H(L)$ as a function of $L$ in Fig.~\ref{fig:ph000}, where the error bar is governed
 by performing a numerical differentiation of Eq.~\eqref{eq:chi} with the use of the discretized $\kappa$. We observe a clear logarithmic $L$ dependence of $H(L)$, which results in $\alpha\simeq 0$.
The solid line represents a linear fit as a function of $\ln L$, which describes the data very well for a wide range of $L\in [32,1024]$. 
According to the Josephson law, $\alpha$ is related to the critical exponent for the correlation
  length $\nu$ as
  \begin{equation}
    d\,\nu=2-\alpha,
    \label{}
  \end{equation}
which tells us $\nu\simeq 1$. One can also estimate $\nu$ from the finite size scaling behavior
  of the peak position $\kappa_c(L)$:
  \begin{equation}
    \kappa_c(L)-\kappa_c(\infty)\propto L^{-1/\nu}.
    \label{eq:fsr}
  \end{equation}
  Figure~\ref{fig:kc000} shows $L^{-1}$ dependence of $\kappa_c(L)$. 
The solid curve represents the fit result obtained with the fit function of $\kappa_c(L)=\kappa_c(\infty)+a_c L^{-1/\nu}$.
The fit range is chosen as $L\in[64\sqrt{2},1024]$ avoiding possible finite size effects expected in the range of small $L$.  
Numerical values for the fit results are presented in Table~\ref{tab:kc}. The value of $\nu$
  is consistent with $\nu=1$ within the error bar, though its magnitude is rather large.

  \begin{table}
    \centering
    \caption{Results for the finite size scaling analysis on the peak position of the chiral susceptibility.}
    \label{tab:kc}
    \begin{ruledtabular}
    \begin{tabular}{ccccc}\toprule
      $\beta$&$\nu$&$\kappa_c$&fit range&$\chi^2/{\rm d.o.f}$\\ \hline
      $0.0$&$1.24(40)$&$0.380665(59)$&$L\in [64\sqrt{2},1024]$&$0.018$ \\
      $5.0$&$1.01(21)$&$0.27972(27)$&$L\in [16,128]$&$0.17$ \\
      $10.0$&$0.76(20)$&$0.26892(24)$&$L\in [16\sqrt{2},128]$&$0.018$ \\ \bottomrule
    \end{tabular}
    \end{ruledtabular}
  \end{table}

\begin{figure}
    \centering
    \includegraphics[width=80mm]{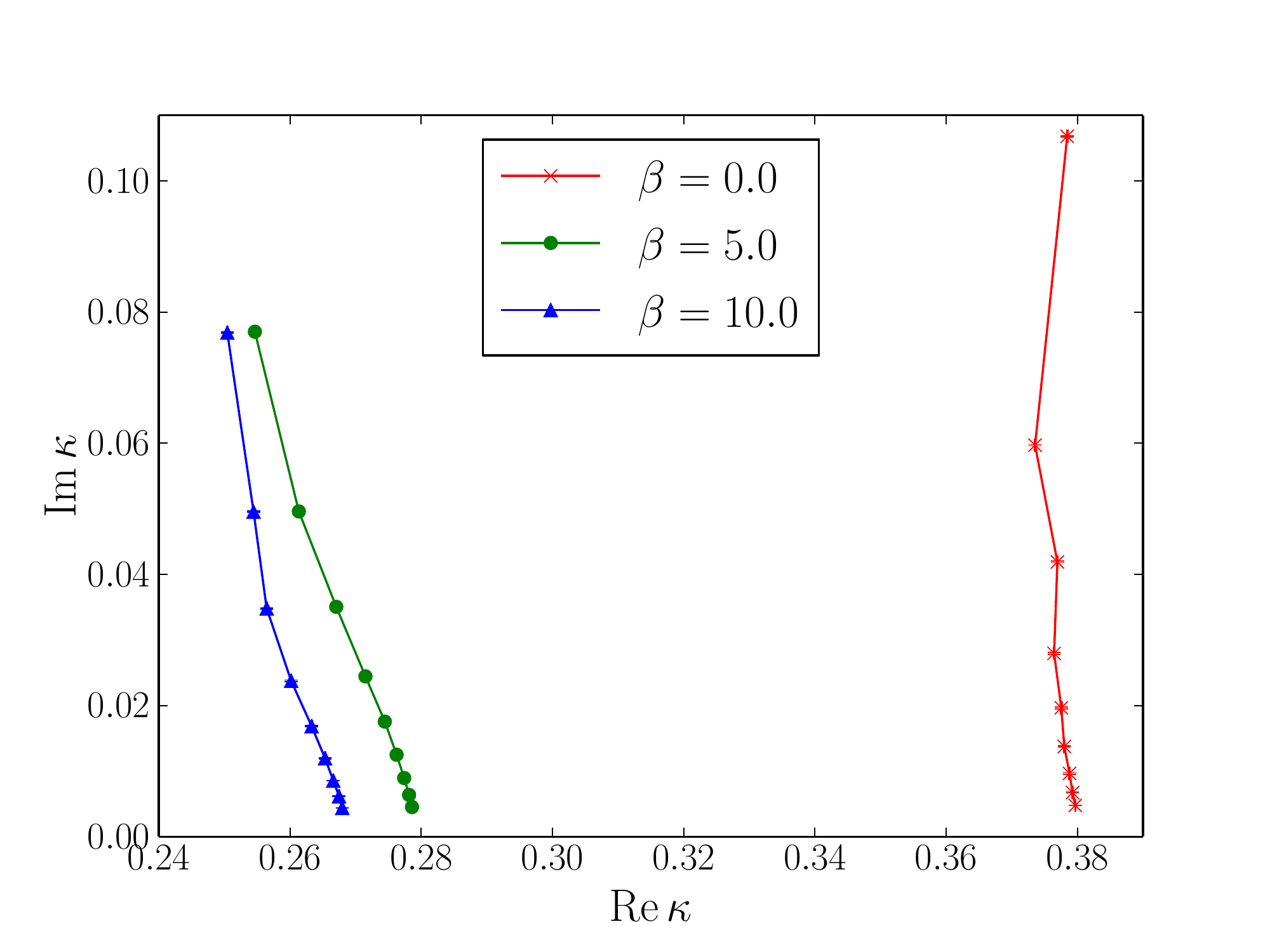}
    \caption{Partition function zeros closest to the real axis in the complex $\kappa$ plane for $L\in[4,64]$.}
    \label{fig:ly}
  \end{figure}

\begin{figure}
    \centering
    \includegraphics[width=80mm]{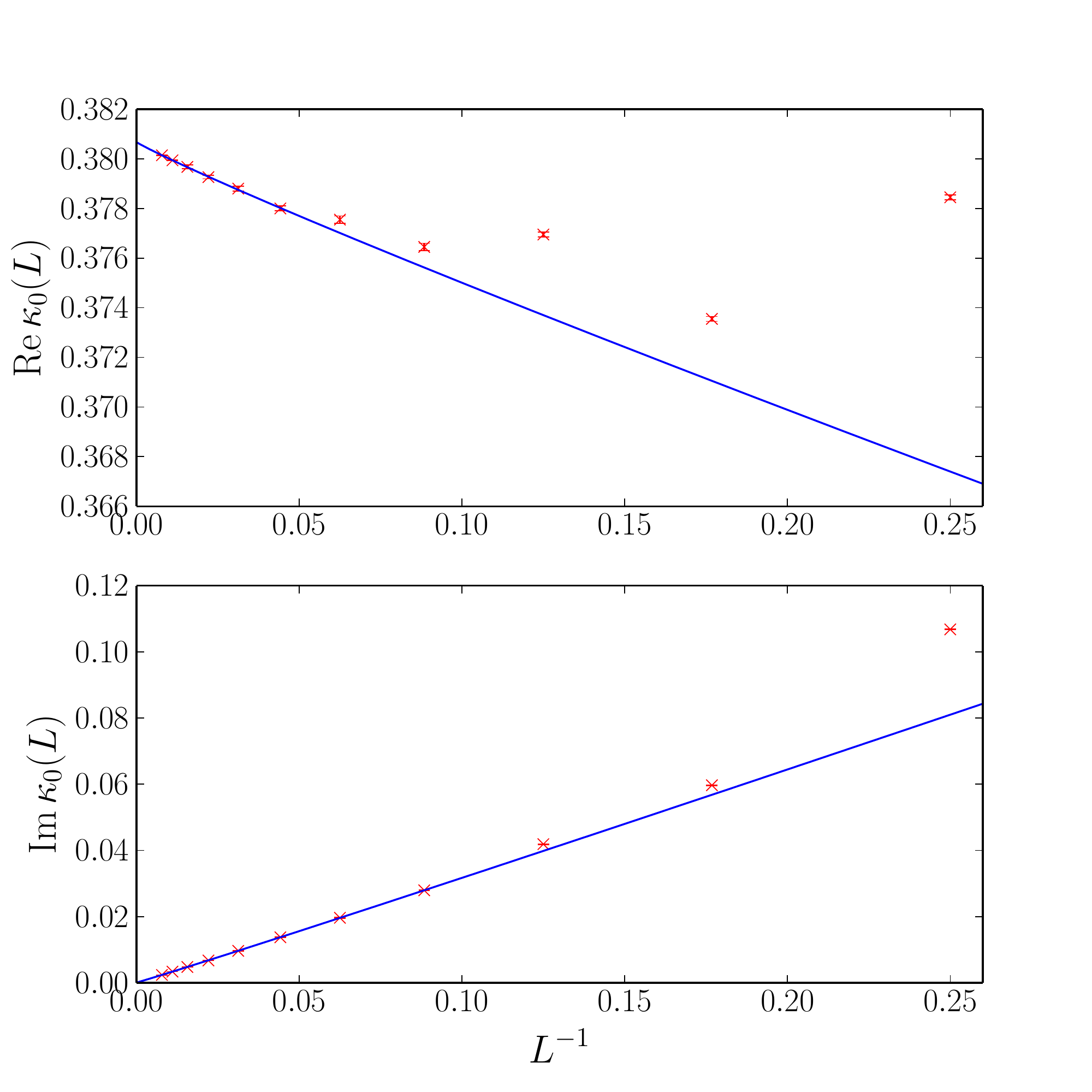}
    \caption{Real (top) and imaginary (bottom) parts of the Lee-Yang zero as a function of $L^{-1}$ at $\beta=0.0$. Solid curves represent the fit results.}
    \label{fig:ly000}
  \end{figure}

  The Lee-Yang zero analysis allows us to determine the critical exponent $\nu$ more accurately.
  Figure~\ref{fig:ly} shows the position of the partition function zero closest to the real axis for $L\in[4,64]$. We refer to it as $\kappa_0(L)$ hereafter.
The $\kappa_0(L)$ is located on the mesh of the discretized $\text{Re}\,\kappa$ and $\text{Im}\,\kappa$ so that the mesh spacing determines the error bars of $\text{Re}\,\kappa_0(L)$ and $\text{Im}\,\kappa_0(L)$. 
  We expect that both the real part and imaginary one of $\kappa_0(L)$ scale to Eq.~\eqref{eq:fsr}:
  \begin{gather}
    \text{Re}\,\kappa_0(L)-\text{Re}\,\kappa_0(\infty)\propto L^{-1/\nu},\label{eq:rek}\\
    \text{Im}\,\kappa_0(L)-\text{Im}\,\kappa_0(\infty)\propto L^{-1/\nu},\label{eq:imk}
  \end{gather}
  where $\text{Re}\,\kappa_0(\infty)=\kappa_c(\infty)$ and $\text{Im}\,\kappa_0(\infty)=0$ should be realized.
  In Fig.~\ref{fig:ly000} we present $\text{Re}\,\kappa_0(L)$ and $\text{Im}\,\kappa_0(L)$ as a function of $L^{-1}$. We observe that $\text{Re}\,\kappa_0(L)$ has 
rather large finite size corrections in smaller $L$ compared 
to $\text{Im}\,\kappa_0(L)$. The solid curves denote
  the fit results with $\text{Re}/\text{Im}\,\kappa_0(L)=\text{Re}/\text{Im}\,\kappa_0(\infty)+a_{R/I} L^{-1/\nu}$ based on Eqs.~\eqref{eq:rek} and \eqref{eq:imk}, whose numerical values are listed in Table~\ref{tab:ly}.
We find that the Lee-Yang zero analysis gives much better precision for the value of $\nu$ than the scaling analysis of the peak position of the chiral susceptibility. This could be expected by comparing the $L^{-1}$ dependence of the peak position of the chiral susceptibility in Fig.~\ref{fig:kc000} and that of the Lee-Yang zero in Fig.~\ref{fig:ly000}: The latter shows better scaling behavior from the smaller $L$. 
  Both results for $\text{Re}\,\kappa_0(L)$ and $\text{Im}\,\kappa_0(L)$ indicate $\nu\simeq 1$. We should also note that $\text{Re}\,\kappa_0(\infty)$ is consistent with $\kappa_c(\infty)$ determined by the peak position of the chiral susceptibility and 
$\text{Im}\,\kappa_0(\infty)$ vanishes in the infinite volume limit as expected.  We conclude that our results at the strong coupling limit indicate a second-order phase transition with $\alpha\simeq 0$ and $\nu=1$  which belongs to the 2d Ising universality class. It should be
  noted that our result for $\kappa_c(\infty)$ is also consistent with $\kappa_c=0.3805(1)$ obtained by the analysis based on an eight-vertex model in Ref.~\cite{Gausterer:1995np}.

  \begin{table}
    \centering
    \caption{Results for the finite size scaling analysis on the real part (top) and the imaginary part
      (bottom) of the Lee-Yang zero.}
    \label{tab:ly}
    \begin{ruledtabular}
    \begin{tabular}{ccccc}\toprule
      $\beta$&$\nu$&$\text{Re}\,\kappa_0(\infty)$&fit range&$\chi^2/{\rm d.o.f}$\\ \hline
      $0.0$&$1.08(10)$&$0.38067(10)$&$L\in [16\sqrt{2},128]$&$0.13$ \\
      $5.0$&$0.765(24)$&$0.27943(10)$&$L\in [8,64]$&$0.086$ \\
      $10.0$&$0.776(39)$&$0.26892(16)$&$L\in [8\sqrt{2},64]$&$0.10$ \\ \bottomrule
    \end{tabular}
    \end{ruledtabular}
    \begin{ruledtabular}
    \begin{tabular}{ccccc}\toprule
      $\beta$&$\nu$&$\text{Im}\,\kappa_0(\infty)$&fit range&$\chi^2/{\rm d.o.f}$\\ \hline
      $0.0$&$0.9755(80)$&$0.000062(52)$&$L\in [8\sqrt{2},128]$&$0.23$ \\
      $5.0$&$0.994(14)$&$0.00031(17)$&$L\in [8,64]$&$2.3$ \\
      $10.0$&$0.995(19)$&$0.00029(20)$&$L\in [8\sqrt{2},64]$&$0.22$ \\ \bottomrule
    \end{tabular}
    \end{ruledtabular}
  \end{table}

  \subsection{Finite coupling ($\beta=5.0,\,10.0$)}
\begin{figure}
    \centering
    \includegraphics[width=80mm]{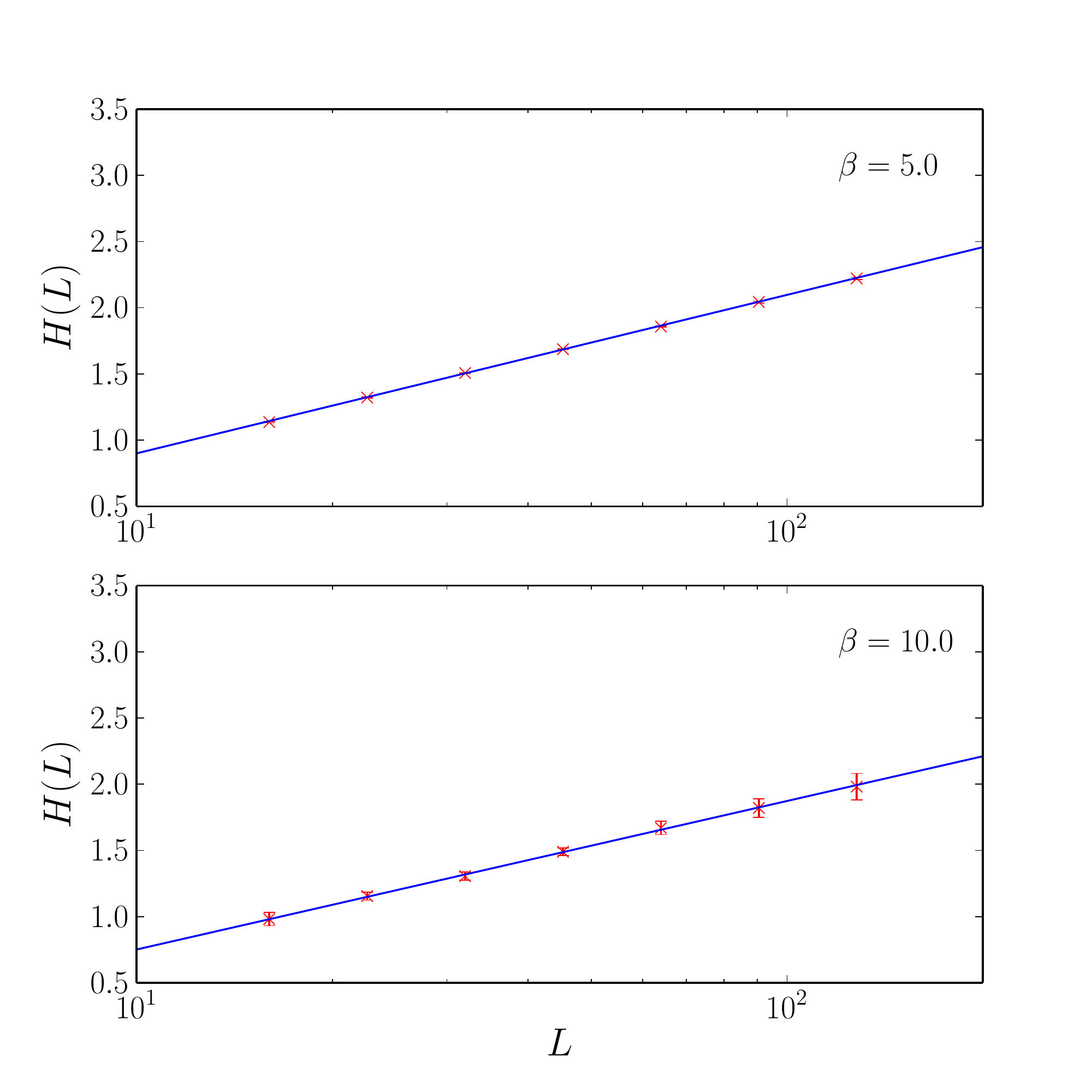}
    \caption{Peak height of the chiral susceptibility $H(L)$ as a function of $L$ at $\beta=5.0$ (top) and $\beta=10.0$
      (bottom). The horizontal axis is logarithmic. Solid lines represent linear fits in terms of $\ln Z$.}
    \label{fig:ph510}
  \end{figure}

\begin{figure}
    \centering
    \includegraphics[width=80mm]{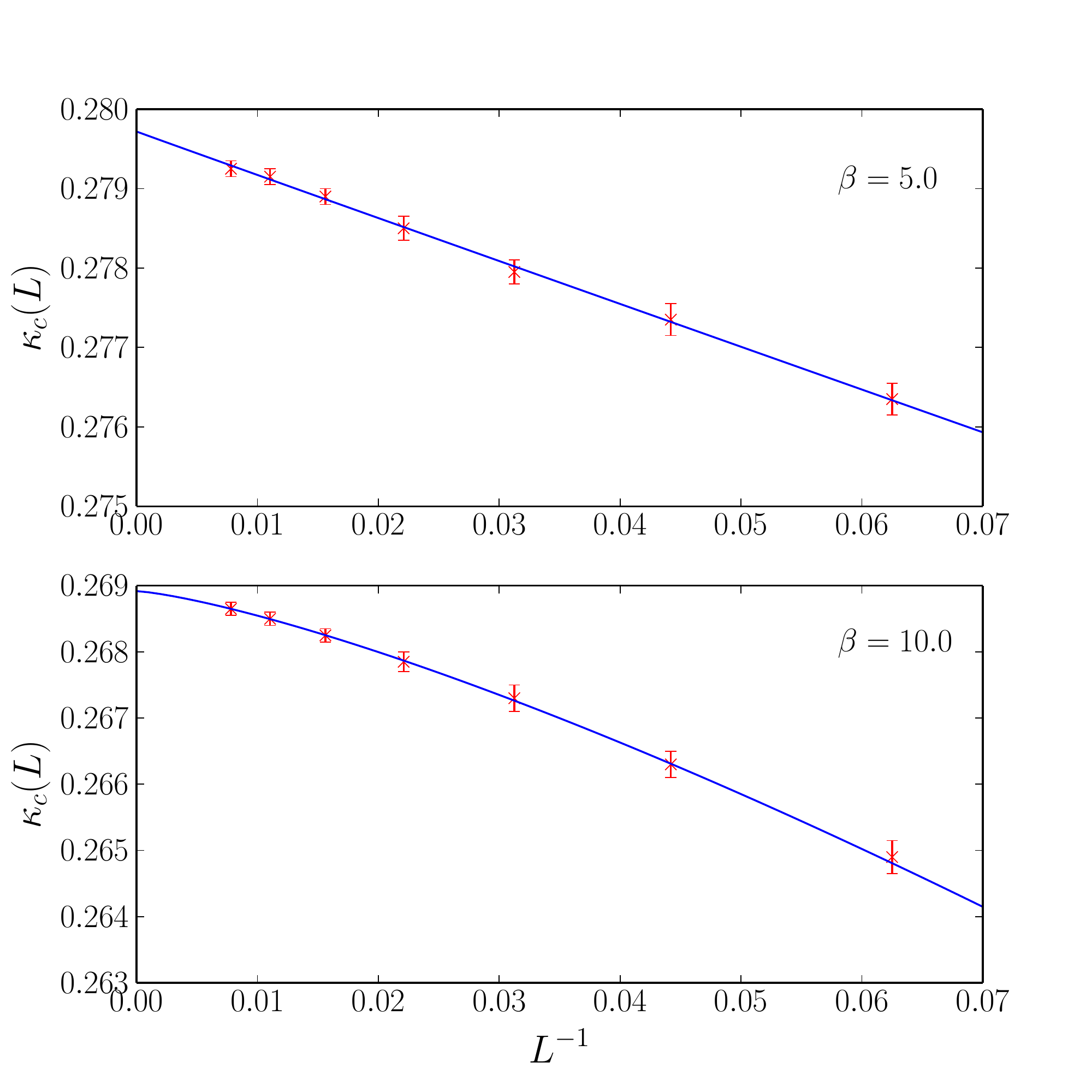}
    \caption{Peak position of the chiral susceptibility $\kappa_c(L)$ as a function of $L^{-1}$ at $\beta=5.0$ (top) and $\beta=10.0$
    (bottom). Solid curves represent the fit results. }
    \label{fig:kc510}
  \end{figure}

  Since the numerical accuracy of the GTRG at finite coupling becomes worse than at the strong coupling limit, we perform finite size scaling analyses on smaller lattices. We first investigate the scaling behavior of the peak height and position for the chiral susceptibility.
  In Fig.~\ref{fig:ph510} we plot $H(L)$ as a function of $L$ at $\beta=5.0$ and $\beta=10.0$ for $L\in[16,128]$. The solid lines denote the linear fits   
in terms of $\ln Z$ for $L\in [32,128]$.
Both plots show clear logarithmic dependence on $L$ as in the strong coupling limit, which indicates $\alpha\simeq 0$.
  We also plot $\kappa_c(L)$ as a function of $L^{-1}$ in Fig.~\ref{fig:kc510}, where the solid curves denote
  the fit results with $\kappa_c(L)=\kappa_c(\infty)+a_c L^{-1/\nu}$ based on Eq.~\eqref{eq:fsr}.
The fit range is chosen as $L\in[16,128]$ at $\beta=5.0$ and $L\in[16\sqrt{2},128]$ at $\beta=10.0$.  
Numerical values for the fit results are summarized in Table~\ref{tab:kc}. The value of $\nu$ indicates consistency with $\nu=1$ taking account of the rather large error bar. As in the strong coupling limit, it is hard to determine the value of $\nu$ with good precision from the scaling behavior of the peak position of the chiral susceptibility.  

\begin{figure}
    \centering
    \includegraphics[width=80mm]{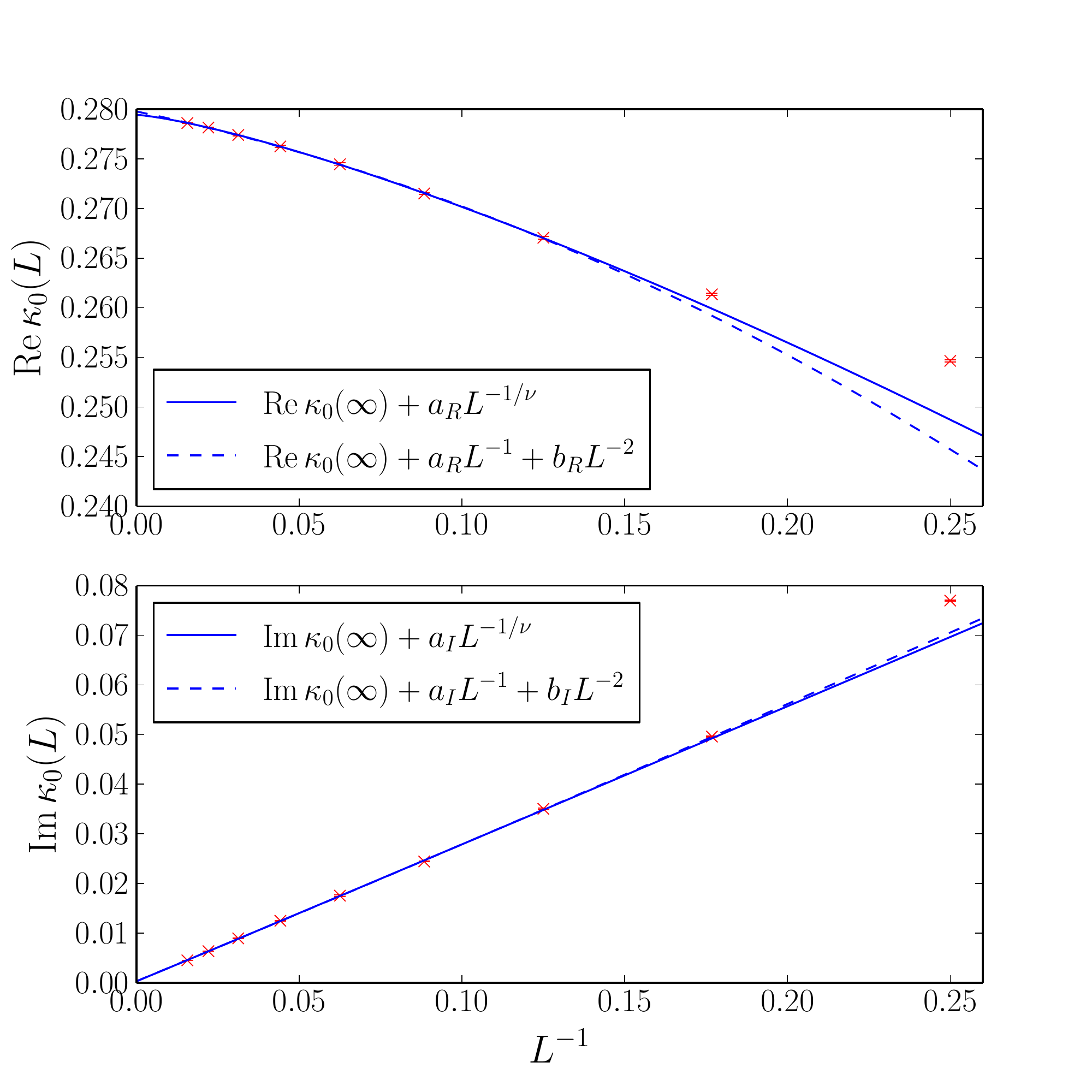}
    \caption{Real (top) and imaginary (bottom) parts of the Lee-Yang zero as a function of $L^{-1}$ at $\beta=5.0$. Solid curves represent the fit results with $\text{Re}/\text{Im}\,\kappa_0(L)=\text{Re}/\text{Im}\,\kappa_0(\infty)+a_{R/I} L^{-1/\nu}$ and dotted ones with Eqs.~\eqref{eq:jfre} and \eqref{eq:jfim}.}
    \label{fig:ly050}
  \end{figure}

\begin{figure}
    \centering
    \includegraphics[width=80mm]{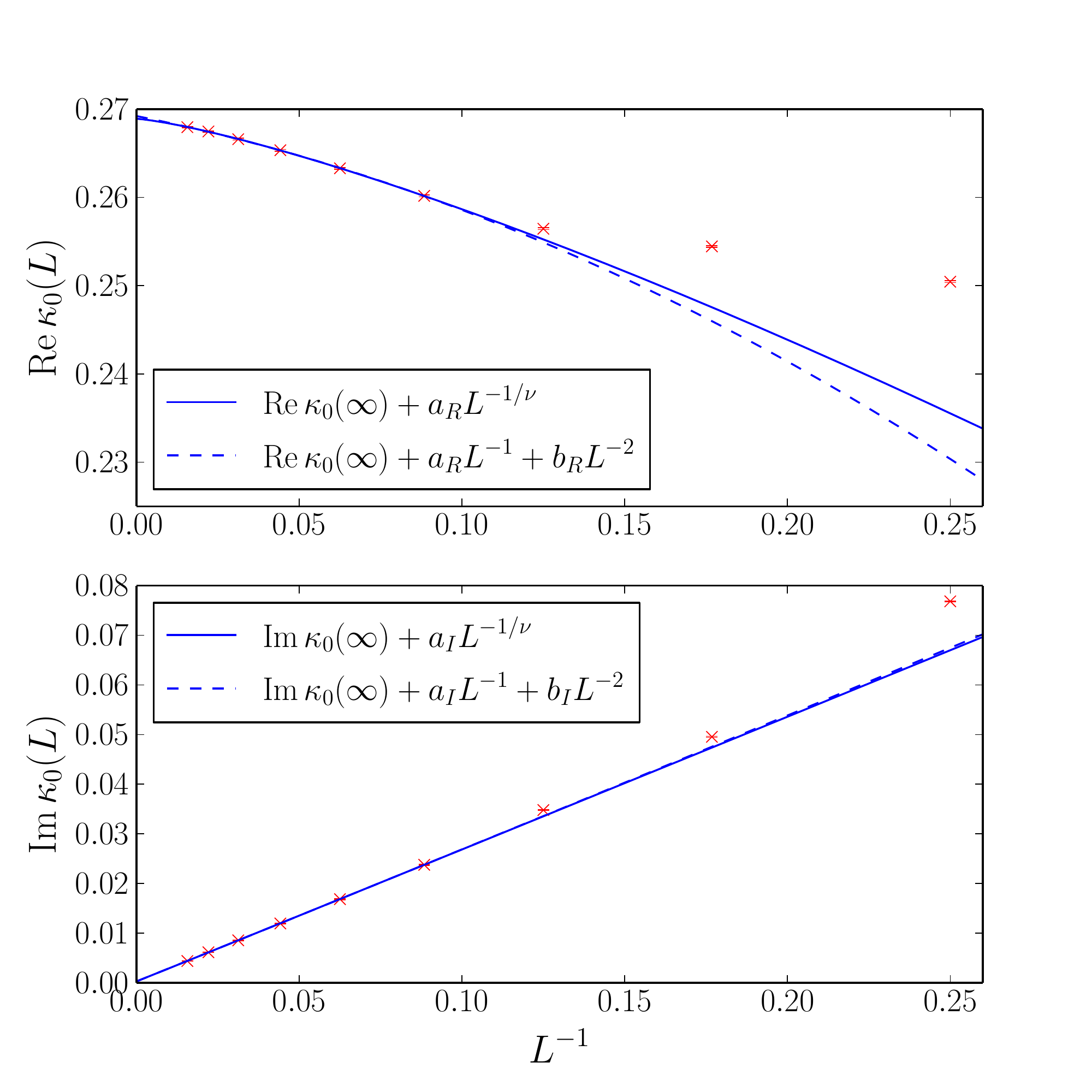}
    \caption{Real (top) and imaginary (bottom) parts of the Lee-Yang zero as a function of $L^{-1}$ at $\beta=10.0$. Solid curves represent the fit results with $\text{Re}/\text{Im}\,\kappa_0(L)=\text{Re}/\text{Im}\,\kappa_0(\infty)+a_{R/I} L^{-1/\nu}$ and dotted ones with Eqs.~\eqref{eq:jfre} and \eqref{eq:jfim}.}
    \label{fig:ly100}
  \end{figure}

  In the strong coupling limit we know that a more accurate evaluation of $\nu$ is obtained from the Lee-Yang zero analysis.
  Figures~\ref{fig:ly050} and \ref{fig:ly100} show finite size scaling plots of both the real and
  imaginary parts of the Lee-Yang zero at $\beta=5.0$ and 10.0, respectively.
We employ the same fit procedure as in the strong coupling limit.
Numerical values of the fit results are given in Table~\ref{tab:ly} together with the fit ranges.
  While the results for the imaginary part indicate $\nu=1$ with very good precision, those for the real part show disagreement with $\nu=1$ beyond the error bars. A similar inconsistency is reported in Ref.~\cite{Gausterer:1995np}, where the authors
  argue that the real part of the Lee-Yang zero has little chance to exhibit the leading scaling
  behaviour because it changes very little as the lattice size $L$ increases. 
The same features are observed in our results of Fig.~\ref{fig:ly}.
  We also investigate possible finite size contaminations in $\text{Re}\,\kappa_0(L)$ and $\text{Im}\,\kappa_0(L)$ by employing the following fit functions:
  \begin{gather}
    \text{Re}\,\kappa_0(L)-\text{Re}\,\kappa_0(\infty)=a_RL^{-1}+b_RL^{-2},
    \label{eq:jfre}\\
    \text{Im}\,\kappa_0(L)-\text{Im}\,\kappa_0(\infty)=a_IL^{-1}+b_IL^{-2},
    \label{eq:jfim}
  \end{gather}
where we assume $\nu=1$ and the $L^{-2}$ term represents the sub-leading contribution. The fit results are depicted with the dotted curves in Figs.~\ref{fig:ly050} and \ref{fig:ly100} and the numerical values for the coefficients $a_{R/I}$ and $b_{R/I}$ are presented in Table~\ref{tab:jf}.
  We find that the coefficient $b_R$ has much larger magnitude than $a_R$, which results in large $L^{-2}$
  contributions to $\text{Re}\,\kappa_0(L)$. On the other hand, the imaginary part shows that the coefficient $b_I$ is negligibly small compared to $a_I$. This assures us that the Lee-Yang zero analysis of the imaginary part is more reliable than the real one avoiding the possible sub-leading contaminations. The similar situation is also found with a different choice of the boundary condition in Ref.~\cite{Gausterer:1995np}.
  In conclusion, our results indicate a second-order phase transition with $\alpha\simeq 0$ and $\nu=1$ so that the one-flavor lattice Schwinger model belongs to the 2d Ising universality
  class even at finite coupling. This agrees with the result obtained by neither the microcanonical
  fermionic average approach~\cite{Azcoiti:1995bx} nor the weak coupling expansion~\cite{Kenna:1998rs}.

  \begin{table}
    \centering
    \caption{Fit results including the sub-leading finite size contribution. The fit ranges are the same as in Table~\ref{tab:ly}.}
    \label{tab:jf}
    \begin{ruledtabular}
    \begin{tabular}{ccccc}\toprule
      $\beta$&$\text{Re}\,\kappa_0(\infty)$&$a_R$&$b_R$&$\chi^2/{\rm d.o.f}$\\ \hline
      $5.0$&$0.279773(91)$&$-0.0687(45)$&$-0.270(36)$&$0.55$ \\
      $10.0$&$0.26922(14)$&$-0.0736(73)$&$-0.327(72)$&$0.25$ \\ \bottomrule
    \end{tabular}
    \begin{tabular}{ccccc}\toprule
      $\beta$&$\text{Im}\,\kappa_0(\infty)$&$a_I$&$b_I$&$\chi^2/{\rm d.o.f}$\\ \hline
      $5.0$&$0.00032(13)$&$0.2722(66)$&$0.035(53)$&$2.1$ \\
      $10.0$&$0.00028(11)$&$0.2641(60)$&$0.019(61)$&$0.21$ \\ \bottomrule
    \end{tabular}
    \end{ruledtabular}
  \end{table}

  \section{Summary and Outlook}
  \label{sec:sum}
  We have applied the GTRG to the one-flavor lattice Schwinger model with the Wilson fermion formulation.
  The finite size scaling analyses of the peak height of the chiral susceptibility and the Lee-Yang zero
  show that the phase transition not only at the strong coupling limit but also at finite coupling belongs to
  the same universality class as the 2d Ising model similarly to the free fermion case. It tells us that we
  can take the massless continuum limit along the critical line $\kappa=\kappa_c(\beta)$.

  This is the first application of the GTRG to lattice gauge theory including fermions. The GTRG has a strong advantage that it does not
  suffer from the sign problem caused by the fermion determinant, which is demonstrated in this work.  A further possibility is an application of the GTRG to the physical system with the $\theta$ term where the action is a complex number.
  A numerical analysis of the lattice Schwinger model with the $\theta$ term is under way.

  There remain some difficulties in extending the GTRG to lattice QCD.
  Although our method can be formally extended to 2d lattice QCD by adopting a tensor network formulation of
  SU($N$) gauge theory  proposed by Liu {\it et al.}~\cite{Liu:2013nsa}, 
it is necessary to check how much computational cost is actually required for numerical calculations.
  The biggest difficulty is to develop a practical method to calculate 4d systems.
  The HOTRG~\cite{Xie:2012aa} which is based on the higher-order SVD instead of the matrix SVD is the most
  effective approach to higher dimensional systems at the moment.
  Its computational cost, however, is proportional 
to $D^{15}$ for a 4d hypercubic lattice, which is still too expensive.

  \begin{acknowledgments}
    We would like to thank Y. Nakamura for his helpful advice.
    This work is supported by the Large Scale Simulation Program No.~T12/13-01 (FY2012-13)
    of High Energy Accelerator Research Organization (KEK).
    Part of the calculations were performed on the cluster systems at RIKEN Advanced Institute
    for Computational Science.
  \end{acknowledgments}

  \appendix
  \section{Values for $W^{i_{f1},i_{f2},j_{f1},j_{f2},k_{f1},k_{f2},l_{f1},l_{f2}}$ and $H_{m_b,n_b}^{i_{f1},i_{f2}}$}
  \label{sec:appa}
  We present the values for $W^{i_{f1},i_{f2},j_{f1},j_{f2},k_{f1},k_{f2},l_{f1},l_{f2}}$ and $H_{m_b,n_b}^{i_{f1},i_{f2}}$ introduced in Sec.~\ref{sec:gtn}:
  \begin{gather}
    \renewcommand{\arraystretch}{1.5}
    \begin{array}{ll}
      W^{0,0,0,0,0,0,0,0}=\left( \frac{1}{2\kappa} \right)^2,&
      W^{1,0,0,0,1,0,0,0}=\frac{1}{2\kappa},\\
      W^{0,1,0,0,0,1,0,0}=-\frac{1}{2\kappa},&
      W^{1,1,0,0,1,1,0,0}=1,\\
      W^{0,0,1,0,0,0,1,0}=\frac{1}{2\kappa},&
      W^{0,0,0,1,0,0,0,1}=-\frac{1}{2\kappa},\\
      W^{0,0,1,1,0,0,1,1}=1,&
      W^{1,0,0,1,0,1,1,0}=1,\\
      W^{0,1,1,0,1,0,0,1}=1,&
      W^{1,0,1,0,1,0,1,0}=-\frac{1}{2},\\
      W^{1,0,0,1,1,0,0,1}=\frac{1}{2},&
      W^{0,1,1,0,0,1,1,0}=\frac{1}{2},\\
      W^{0,1,0,1,0,1,0,1}=-\frac{1}{2},&
      W^{1,1,1,1,0,0,0,0}=-\frac{1}{2},\\
      W^{1,1,0,0,0,0,1,1}=-\frac{1}{2},&
      W^{0,0,1,1,1,1,0,0}=-\frac{1}{2},\\
      W^{0,0,0,0,1,1,1,1}=-\frac{1}{2},&
      W^{1,1,1,0,0,0,1,0}=\frac{1}{2},\\
      W^{1,1,0,1,0,0,0,1}=\frac{1}{2},&
      W^{1,0,1,1,1,0,0,0}=\frac{1}{2},\\
      W^{0,1,1,1,0,1,0,0}=\frac{1}{2},&
      W^{0,0,1,0,1,1,1,0}=-\frac{1}{2},\\
      W^{0,0,0,1,1,1,0,1}=-\frac{1}{2},&
      W^{1,0,0,0,1,0,1,1}=-\frac{1}{2},\\
      W^{0,1,0,0,0,1,1,1}=-\frac{1}{2},&
      W^{1,1,1,0,1,0,0,0}=\frac{1}{\sqrt{2}},\\
      W^{1,1,0,1,0,1,0,0}=-\frac{1}{\sqrt{2}},&
      W^{1,1,0,0,1,0,0,1}=\frac{1}{\sqrt{2}},\\
      W^{1,1,0,0,0,1,1,0}=\frac{1}{\sqrt{2}},&
      W^{1,0,1,1,0,0,1,0}=\frac{1}{\sqrt{2}},\\
      W^{0,1,1,1,0,0,0,1}=-\frac{1}{\sqrt{2}},&
      W^{0,0,1,1,1,0,0,1}=-\frac{1}{\sqrt{2}},\\
      W^{0,0,1,1,0,1,1,0}=-\frac{1}{\sqrt{2}},&
      W^{1,0,0,1,1,1,0,0}=\frac{1}{\sqrt{2}},
    \end{array}\displaybreak[3]\nonumber\\
    \renewcommand{\arraystretch}{1.5}
    \begin{array}{ll}
      W^{1,0,0,0,1,1,1,0}=-\frac{1}{\sqrt{2}},&
      W^{0,1,1,0,1,1,0,0}=\frac{1}{\sqrt{2}},\\
      W^{0,1,0,0,1,1,0,1}=\frac{1}{\sqrt{2}},&
      W^{1,0,0,1,0,0,1,1}=-\frac{1}{\sqrt{2}},\\
      W^{0,1,1,0,0,0,1,1}=-\frac{1}{\sqrt{2}},&
      W^{0,0,1,0,1,0,1,1}=-\frac{1}{\sqrt{2}},\\
      W^{0,0,0,1,0,1,1,1}=\frac{1}{\sqrt{2}},&
      W^{1,0,0,1,0,0,0,0}=\frac{1}{2\sqrt{2}\kappa},\\
      W^{1,0,0,0,0,0,1,0}=\frac{1}{2\sqrt{2}\kappa},&
      W^{0,1,1,0,0,0,0,0}=-\frac{1}{2\sqrt{2}\kappa},\\
      W^{0,1,0,0,0,0,0,1}=\frac{1}{2\sqrt{2}\kappa},&
      W^{0,0,1,0,1,0,0,0}=\frac{1}{2\sqrt{2}\kappa},\\
      W^{0,0,0,1,0,1,0,0}=\frac{1}{2\sqrt{2}\kappa},&
      W^{0,0,0,0,1,0,0,1}=-\frac{1}{2\sqrt{2}\kappa},
    \end{array}\displaybreak[4]\nonumber\\
    \renewcommand{\arraystretch}{1.5}
    \begin{array}{ll}
      W^{0,0,0,0,0,1,1,0}=\frac{1}{2\sqrt{2}\kappa},&
      \text{others}=0,
    \end{array}\displaybreak[0]\\
    \renewcommand{\arraystretch}{1.5}
    \begin{array}{c}
      H_{m_b,n_b}^{0,0}=H_{m_b,n_b}^{1,1}=\delta_{m_b,n_b},\displaybreak[3]\\
      H_{m_b,n_b}^{1,0}=-\delta_{m_b,n_b+1},\displaybreak[3]\\
      H_{m_b,n_b}^{0,1}=\delta_{m_b,n_b-1},
    \end{array}
  \end{gather}
  where $W^{i_{f1},i_{f2},j_{f1},j_{f2},k_{f1},k_{f2},l_{f1},l_{f2}}$ has only 49 non-zero components.

\end{document}